\input{aipcheck}

\documentclass[
    ,final            % use final for the camera ready runs
  ]
  {aipproc}

\layoutstyle{8x11double}
\usepackage{bm}
\begin{document}

\title{Quasistatic behavior and force transmission in packing of irregular polyhedral particles}


\author{Emilien Az\'ema}{
  address={LMGC, Universit\'e Montpellier 2, 34080 Montpellier cedex 05, France}
}

\author{Farhang. Radjai}{
  address={LMGC, Universit\'e Montpellier 2, 34080 Montpellier cedex 05, France}
}

\author{Gilles Saussine}{
  address={Innovation and Research Departement of SNCF, 45 rue de Londres, 75379 Paris Cedex 08, France}
}

\begin{abstract}
Dense packings composed of irregular polyhedral particles are investigated by numerical 
simulations under quasistatic triaxial compression. The Contact Dynamics method is used
for this investigation with 40 000 particles. The effect of particle shape is analyzed by comparing 
this packing with a packing of similar particle size distribution but with spherical particles.
We analyze the origin of the higher shear strength 
of the polyhedra packing by considering various anisotropy parameters characterizing the 
microstructure and force transmission. Remarkably, we find that the polyhedra packing has a 
lower fabric anisotropy in terms of branch vectors (joining the particle centers) than the sphere 
packing. In contrast, the polyhedra packing shows a much higher force anisotropy which is at the 
origin of its higher shear strength. The force anisotropy in the polyhedra packing is shown to be 
related to the formation of face-face contacts. In particular, most face-face contacts 
belong to strong force chains along the major principal stress direction whereas vertex-face and 
edge-edge contacts are correlated with weak forces and oriented on average along the minor 
principal stress direction in steady shearing.

\end{abstract}

\maketitle

%--------------------------------------------------------------------------------------------------
\section{Introduction}
During the last two decades, granular media  composed of 
circular particles (in 2D) and  spheres (in 3D) have been a subject of 
systematic research. In particular, various microscopic features such as 
fabric anisotropy \cite{Rothenburg1989}, 
force transmission \cite{Radjai1996,Coppersmith1996,Mueth1998a,Silbert2002,Majmudar2005}
and friction mobilization \cite{Radjai1998}
have been analyzed numerically and experimentally.
Hence, an emerging issue today is how 
robust these findings are with respect to particle properties such as shape and 
size distribution  
\cite{Ouadfel2001,Antony2004,Azema2007}.  

In this paper, we study numerically granular materials 
composed of polyhedral particles. We use the contact dynamics method to 
simulate the slow shear of these materials in comparison to sphere packings 
with similar particle size distribution. 
The facetted shapes give rise to a rich microstructure where the particles 
touch at their faces, edges and vertices. We analyze the fabric and force anisotropies 
and their link with the stress-strain behavior. 
We show that face-face contacts  
play a major role in force transmission and statics of polyhedra by accommodating  
long force chains that are basically unstable in a packing composed 
of spheres. 

\section{Numerical procedures}
The simulations were carried out by means of the contact dynamics (CD) method with 
irregular polyhedra particles \cite{Jean1999,Moreau1994}. 
We used LMGC90 which is a multipurpose software  
developed in our laboratory, capable of modeling a collection of deformable or undeformable particles of
various shapes by different algorithms \cite{DUBOIS2003}.

\begin{figure}
\centering
\includegraphics[width=0.5\columnwidth]{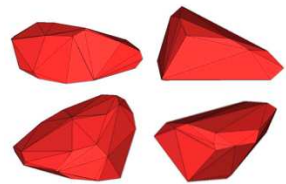}
\caption{Examples of polyhedra used in the simulations.}
\label{grains_ballast}    
\end{figure}

\begin{figure}
\includegraphics[width=0.8\columnwidth]{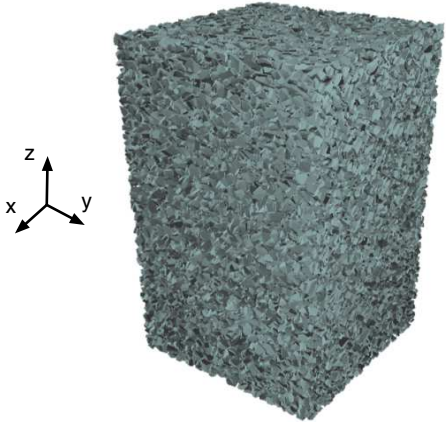}
\caption{A snapshot of the packing S1 (polyhedra). The walls are not shown.}
\label{SamplesPOLYH_SPHER}    
\end{figure}

We generate two numerical samples. The first sample (S1) is composed of 
36933 polyhedra. The particle shapes are taken from a library of 1000 
digitalized ballast grains provided by the French Railway
Company SNCF.
Fig. \ref{grains_ballast} shows several examples 
of the polyhedral particles used in the simulations.   
We used the following size distribution:  
50\% of diameter $d_{min}=2.5$ cm, 34\% of diameter $3.75$ cm, 
16\% of diameter $d_{max}=5$ cm, where $d_{min}$ is defined as two 
times the largest distance between the 
barycenter and the vertices of the particle. 
The second sample (S2) is composed of 19998 spheres 
with exactly the same 
size distribution as in S1. Fig. \ref{SamplesPOLYH_SPHER} shows a 
snapshot of the sample S1 in equilibrium state after deposition and isotropic compression 
under a constant stress of $\sigma_0=10^4$ Pa  in a rectangular box at zero gravity. 

The coefficient of friction is 0.5 between the particles and 0 
with the walls.   
The initial value of the solid fraction is $\rho \simeq 0.6$ in both samples. 
Both samples have a nearly square bottom of side such that $L\approx l$ 
and an aspect ratio $H/L \simeq 2$, where $H$ is the height. 
The isotropic samples were subjected to vertical compression by downward 
displacement of the top wall at a constant velocity of $1$ cm/s for 
a constant confining stress $\sigma_0$ acting on the lateral walls.

%--------------------------------------------------------------------------------------------------
\section{Shear strength}
In this section, we compare the stress-strain behavior 
between the packings of polyhedra (packing S1) and spheres (packing S2). 
For the estimation of the stress tensor, we use the internal  
moment tensor ${\bm M}^i$ of each particle i 
defined by \cite{Moreau1997}:
\begin{equation}
M^i_{\alpha \beta} = \sum_{c \in i} f_{\alpha}^c r_{\beta}^c,
\label{eq:M}
\end{equation}
where  $f_{\alpha}^c$ is the $\alpha$ component of the force exerted on 
particle i at the contact c, $r_{\beta}^c$ is the $\beta$ component 
of the position vector of the same contact c, and the summation 
runs over all contact neighbors of particle i.

It can be shown that the internal moment of a collection of 
rigid particles is the sum of the internal moments of 
individual particles \cite{Moreau1997}. 
The stress tensor ${\bm \sigma}$ for a packing of volume $V$  
is simply given by  
\begin{equation}
{\bm \sigma } = \frac{1}{V} \sum_{i \in V} {\bm M}^i =   \frac{1}{V}  \sum_{c \in V} f_{\alpha}^c \ell_{\beta}^c, 
\label{eq:M}
\end{equation}
where ${\bm \ell}^c$ is the branch vector joining the centers of 
the two touching particles at the contact $c$. 

Under triaxial conditions with vertical compression, we have 
$\sigma_1 \geq \sigma_2=\sigma_3$, where the $\sigma_\alpha$ are 
the stress principal values.  
We extract the mean stress $p=(\sigma_1 + \sigma_2 + \sigma_3)/3$, 
and the stress deviator $q=(\sigma_1-\sigma_3)/3$.
For our system of perfectly rigid particles,  the stress state is characterized by 
the mean stress $p$ and the normalized shear stress $q/p$.    

The cumulative strain components $\varepsilon_\alpha$ are defined by 
\begin{equation}
\centering
\varepsilon_1=\int_{H_0}^H \frac{dH'}{H'},
\varepsilon_2=\int_{L_0}^L \frac{dL'}{L'},
\varepsilon_3=\int_{l_0}^l \frac{dl'}{l'},
\label{eq:vara}
\end{equation}
where $H_0$, $l_0$ and $L_0$ are the initial height, width and length of the 
simulation box, respectively, and   $\Delta H = H_0 - H$,  $\Delta l = l_0 - l$ 
and  $\Delta L = L_0 - L$ are the corresponding cumulative displacements. 
The cumulative shear strain is defined by $\varepsilon_q \equiv \varepsilon_1-\varepsilon_3$.

\begin{figure}
\centering
\includegraphics[width= 0.8\columnwidth]{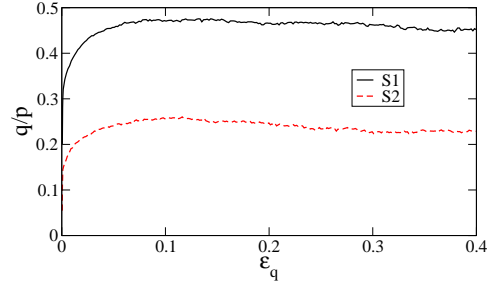}
\caption{The normalized shear stress $q/p$ as a function of shear strain $\varepsilon_q$
for the polyhedra packing S1 and sphere packing S2.}
\label{qp_all_3D}    
\end{figure}

Figure \ref{qp_all_3D} displays the evolution of $q/p$ for the packings S1 and S2 
as a function of $\varepsilon_q$. 
For both packings, we observe a classical behavior characterized by a 
hardening behavior followed by (slight) softening and a stress 
plateau. 
The higher level of $q/p$  for the polyhedra packing reflects the 
organization of the microstructure and the features of force 
transmission that we analyze below.  

%--------------------------------------------------------------------------------------------------
\section{Geometrical anisotropy}
\begin{figure}
\centering
\includegraphics[width=0.5\columnwidth]{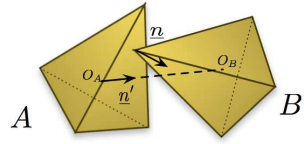}
\caption{Geometry of a contact between two polyhedra.}
\label{image_theta_lambda}   
\end{figure}

For the analyses that will be discussed below, we introduce the local
frame $(\bm n', \bm t')$ where $\bm n'$ is the unit vector along the branch $\bm l$  
and $\bm t'$ is an orthonormal unit vector ; figure \ref{image_theta_lambda}.
We set 
\begin{equation}
\centering
\bm \ell =  \ell \bm n',
\end{equation}
where $\ell$ is the length of the branch vector.

We define the angular averages associated with the branch vectors $\bm \ell$.
Let ${\cal A}(\Omega)$ be the set of  
branch vectors pointing in the direction $\Omega \equiv (\theta,\phi)$ up to a 
solid angle   $d\Omega$ and $N_c(\Omega)$ its cardinal. 
The angular averages are defined as follows:
\begin{equation}
\centering
P_\Omega (\Omega) = \frac{N_c(\Omega)}{N_c},
\; \; \; \;
\langle \ell \rangle (\Omega) = \frac{1}{N_c(\Omega)} \sum_{c\in {\cal A}(\Omega)} \ell^c,
\label{eq:pom}
\end{equation}
where $N_c $ is the total number of contacts, and 
$ \ell^c$,is the actual values of branch vector length, for contact $c$, respectively.  

\begin{figure}
\centering
\includegraphics[width= 0.95\columnwidth]{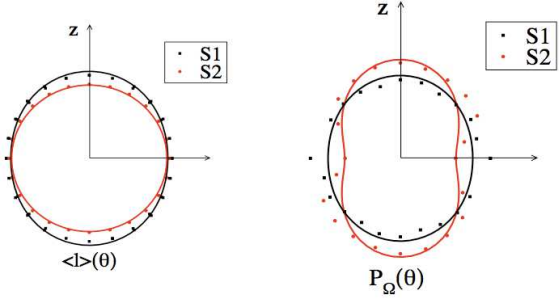}
\caption{Polar representation of the probability density function 
   $P_\Omega$ and $\langle \ell(\Omega) \rangle$ 
   for the samples S1 and S2 in the residual state.}
\label{polar_branche_3D}    
\end{figure}

Under the axisymmetric conditions of our simulations, these two functions 
are independent of $\phi$. Fig. \ref{polar_branche_3D} displays 
a polar representation of these functions in the $\theta$-plane for polyhedra (S1) and 
spheres (S2) at $\varepsilon_q = 0.3$. We observe an anisotropic behavior of 
the unit inter-center vector $P_\Omega(\theta)$ in both cases. A 
weak anisotropy of branch vector can be seen for S1. 
The magnitude of anisotropy is larger for spheres compared to polyhedra 
except for $\langle \ell \rangle (\theta)$ which is weakly anisotropic for polyhedra. 

The simple shapes of the above functions can be approximated by 
 harmonic approximation. Considering only the functions compatible with 
the symmetries of the problem (independent with respect to $\phi$ and 
$\pi$-periodic as a function of $\theta$) we have 
\begin{eqnarray}
\centering
P_ \Omega(\theta) &=& \frac{1}{4\pi} \{ \ 1 + a \ [3\cos^2\theta  -1] \ \}, \label{eq:a1} \\
\langle \ell \rangle (\theta) &=& \ell_0 \{ \ 1 + a_l \ [3\cos^2\theta  -1]  \ \} \label{eq:a2} 
\end{eqnarray}
where $a$, $a_l$ are the anisotropy parameters, $\ell_0$ is the mean branch 
vector length. The probability density function 
$ P_ \Omega(\theta)$ is normalized to 1. The harmonic fits are 
shown in figure \ref{polar_branche_3D} for the
two functions in the critical state.

\begin{figure}[tb]
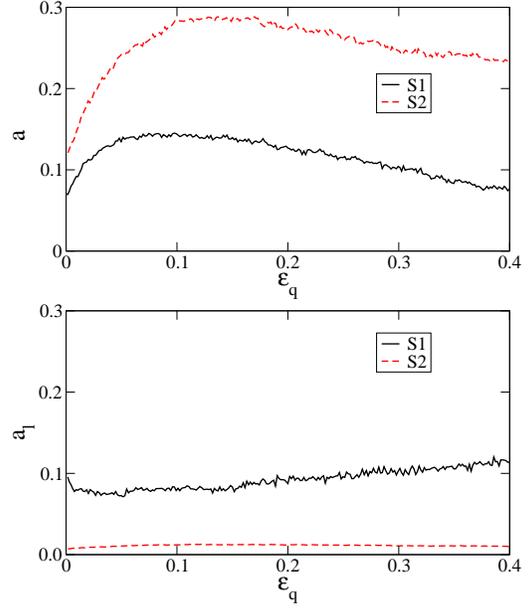
 
  \centering
  \begin{minipage}{0.95\columnwidth}
     \includegraphics[width=0.9\columnwidth]{Figures/fig6a.eps}
     \includegraphics[width= 0.9\columnwidth]{Figures/fig6b.eps}
  \end{minipage}
 \caption{Evolution of anisotropies  $a$, $a_l$
    with $\varepsilon_q$ for packings S1 and S2.}
\label{anisol_epsilon_3D}    
\end{figure}

The evolution of the anisotropies with $\varepsilon_q$ are displayed in 
Fig.  \ref{anisol_epsilon_3D}  for S1 and S2. 
We see that $a$ is systematically larger for spheres than 
for polyhedra. The branch vector length anisotropy $a_l$ is 
negligible for spheres. The low anisotropy of the polyhedra packing results from a particular 
organization of the force network in correlation with the orientations 
of each contacts (edge-to-face, vertex-to-edge...) in 
the packing \cite{Azema2007,Azema2008}. 

%--------------------------------------------------------------------------------------------------
\section{Force anisotropy}

\begin{figure}[tb]
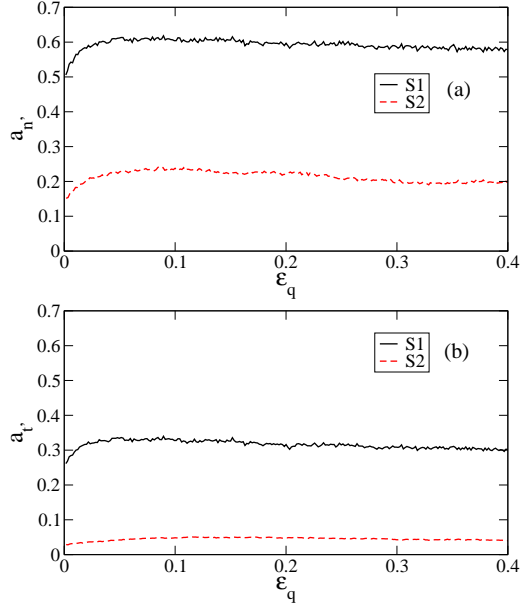
 
  \centering
  \begin{minipage}{0.95\columnwidth}
     \includegraphics[width=0.9\columnwidth]{Figures/fig7c.eps}
     \includegraphics[width=0.9\columnwidth]{Figures/fig7d.eps}
  \end{minipage}
 \caption{Evolution of anisotropies  $a_{n'}$, $a_{t'}$
    with $\varepsilon_q$ for packings S1 and S2.}
\label{anisof_epsilon_3D}    
\end{figure}

We consider the components  of the contact force by  
\begin{equation}
\centering
\bm f =  f_{n'} \bm n' + f_{t'} \bm t'.
\label{f=fn_3D}
\end{equation}
We refer to  $f_{n'}$ and $f_{t'}$ as {\em radial} and {\em orthoradial} 
components of the contact force. 
As the angular orientation of the branch vector, we distinguish 
the angular distributions of radial forces $\langle f_{n'} \rangle (\Omega)$ 
and orthoradial forces $\langle f_{t'} \rangle (\Omega)$.
These two functions can be expanded on a base of spherical harmonics. At leading order, 
we  have \cite{Ouadfel2001,Azema2008} 
\begin{equation}
\left\{
\begin{array}{lcl}
\langle f_{n'}\rangle (\theta) &=& f_0 \{ \ 1 + a_{n'} \ [3\cos^2\theta  -1]  \ \}, \\
\langle f_{t'}\rangle(\theta) &=& f_0 \  a_{t'} \ \sin 2 \theta  , 
\end{array}
\label{eq:a3}
\right.
\label{eqn:fnft}
\end{equation}            
where $a_{n'}$ and $a_{t'}$ are the anisotropy parameters, and $f_0$ the mean force. 
The anisotropies $a_{n'}$ and $a_{t'}$  are plotted in figure \ref{anisof_epsilon_3D} 
as a function of $\varepsilon_q$.
The radial force anisotropy $a_{n'}$ increases as the fabric anisotropy, and tends 
to a plateau. But, in contrast to fabric anisotropy, its value is higher for polyhedra 
than for spheres. This means that the large force anisotropy is correlated 
with particle shape rather than  
with fabric anisotropy. The orthoradial force anisotropy $a_{t'}$ has a 
similar behavior except that it takes considerably higher 
values in the case of polyhedra compared to spheres due to large friction developed by
face to face contacts \cite{Azema2007,Azema2008}.  

\begin{figure}
\centering
\includegraphics[width=0.9\columnwidth]{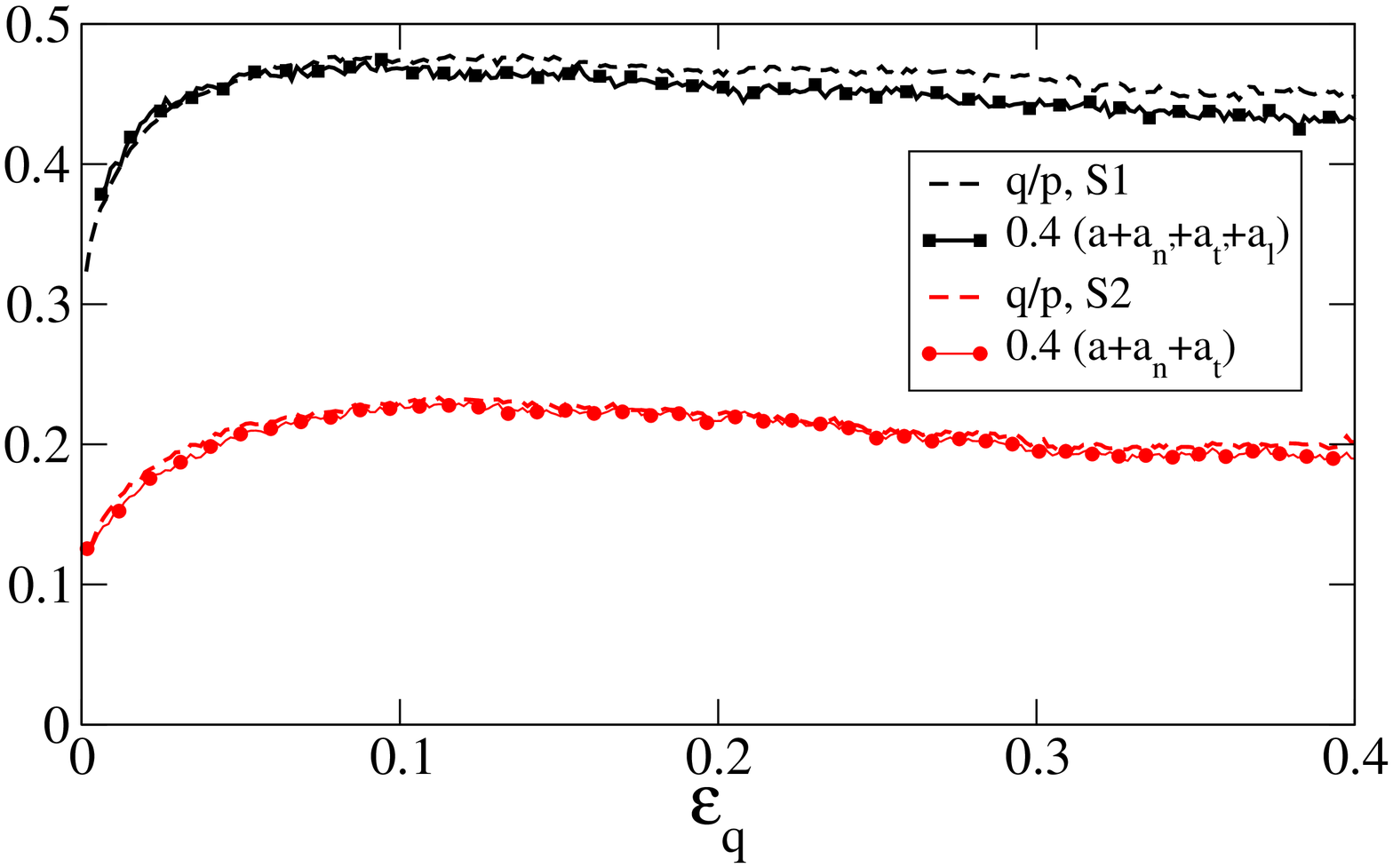}
\caption{The normalized
 shear stress $q/p$ as a function of shear strain $\varepsilon_q$ for the packings S1 and 
 S2 both from direct simulation data and theoretical prediction of  
 Eq. (\ref{eq_qp_3D_general}).}
\label{prediction_qp_3D}    
\end{figure}

The  anisotropies $a$, $a_l$, $a_{n'}$ and $a_{t'}$ are interesting descriptors of granular 
microstructure and force transmission as they underlie the shear stress. Indeed,     
it can be shown that the general expression of the stress tensor Eq. (\ref{eq:M}) under some 
approximations leads to the 
following simple ``stress-force-fabric" relation \cite{Rothenburg1989,Ouadfel2001,Azema2008}:
\begin{equation}     
\frac{q}{p}  \simeq \frac{2}{5} \ (a + a_l+ a_{n'} + a_{t'}),
\label{eq_qp_3D_general}
\end{equation}
As we see in Fig. \ref{prediction_qp_3D}, our simulation data are in 
quantitative agreement with this 
relation both for spheres and polyhedra, all along the shear.  
A remarkable consequence of Eq. (\ref{eq_qp_3D_general}) is to reveal 
that the fabric anisotropy provides a major contribution to  shear stress in the sphere packing whereas 
the force anisotropies are more important for shear stress in 
the polyhedra packing.            

%--------------------------------------------------------------------------------------------------
\section{Conclusion}

The objective of this paper was to isolate the effect of particle shape 
with respect to shear strength in 3D granular media by comparing
two similar packings with different particle shapes. 
A novel finding of this work is 
that the origin of enhanced shear strength in a polyhedra packing compared to a sphere packing 
lies in force anisotropy induced by particle shape. The fabric anisotropy associated with  
the network of branch vectors is lower in the polyhedra packing. 
In other words, 
the force anisotropy, partially underlying  shear strength,  
 is mainly controlled by the fabric anisotropy in a sphere packing. This mechanism 
breaks down to some extent in a packing of polyhedra where 
force anisotropy results mainly from the ``facetted" particle shape.

%--------------------------------------------------------------------------------------------------
%\begin{theacknowledgments}
%\end{theacknowledgments}
\bibliographystyle{aipproc}   
\bibliography{emilien}

\begin{thebibliography}{15}
\expandafter\ifx\csname natexlab\endcsname\relax\def\natexlab#1{#1}\fi
\providecommand{\enquote}[1]{``#1''}
\expandafter\ifx\csname url\endcsname\relax
  \def\url#1{\texttt{#1}}\fi
\expandafter\ifx\csname urlprefix\endcsname\relax\def\urlprefix{URL }\fi
\providecommand{\eprint}[2][]{\url{#2}}

\bibitem[Rothenburg and Bathurst(1989)]{Rothenburg1989}
L.~Rothenburg, and R.~J. Bathurst, \emph{Geotechnique} \textbf{39}, 601--614
  (1989).

\bibitem[Radjai et~al.(1996)]{Radjai1996}
F.~Radjai, M.~Jean, J.~Moreau, and S.~Roux, \emph{Phys. Rev. Letter}
  \textbf{77}, 274--277 (1996).

\bibitem[Coppersmith et~al.(1996)]{Coppersmith1996}
S.~N. Coppersmith, C.~Liu, S.~Majumdar, O.~Narayan, and T.~A. Witten,
  \emph{Phys. Rev. E} \textbf{53}, 4673--4685 (1996).

\bibitem[Mueth et~al.(1998)]{Mueth1998a}
D.~M. Mueth, H.~M. Jaeger, and S.~R. Nagel, \emph{Phys. Rev. E.} \textbf{57},
  3164--3169 (1998).

\bibitem[Silbert et~al.(2002)]{Silbert2002}
L.~E. Silbert, G.~S. Grest, and J.~W. Landry, \emph{Phys. Rev. E} \textbf{66},
  1--9 (2002).

\bibitem[Majmudar and Behringer(2005)]{Majmudar2005}
T.~S. Majmudar, and R.~P. Behringer, \emph{Nature} \textbf{435}, 1079--1082
  (2005).

\bibitem[Radjai et~al.(1998)]{Radjai1998}
F.~Radjai, D.~E. Wolf, M.~Jean, and J.~Moreau, \emph{Phys. Rev. Letter}
  \textbf{80}, 61--64 (1998).

\bibitem[Ouadfel and Rothenburg(2001)]{Ouadfel2001}
H.~Ouadfel, and L.~Rothenburg, \emph{Mechanics of Materials} \textbf{33},
  201--221 (2001).

\bibitem[Antony and Kuhn(2004)]{Antony2004}
S.~Antony, and M.~Kuhn, \emph{International Journal of Solids and Structures}
  \textbf{41}, 5863--5870 (2004).

\bibitem[Az\'ema et~al.(2007)]{Azema2007}
E.~Az\'ema, F.~Radjai, R.~Peyroux, and G.~Saussine, \emph{Phys. Rev. E}
  \textbf{76}, 011301 (2007).

\bibitem[Jean(1999)]{Jean1999}
M.~Jean, \emph{Computer Methods in Applied Mechanic and Engineering}
  \textbf{177}, 235--257 (1999).

\bibitem[Moreau(1994)]{Moreau1994}
J.~Moreau, \emph{Eur. J. Mech. A/Solids} \textbf{13}, 93--114 (1994).

\bibitem[Dubois and Jean(2003)]{DUBOIS2003}
F.~Dubois, and M.~Jean, \enquote{LMGC90 une plateforme de d\'eveloppement
  d\'edi\'ee \`a la mod\'elisation des probl\`emes d'int\'eraction.,} in
  \emph{Actes du sixi\`eme colloque national en calcul des structures -
  CSMA-AFM-LMS -}, 2003, vol.~1, pp. 111--118.

\bibitem[Moreau(1997)]{Moreau1997}
J.~J. Moreau, \enquote{Numerical Investigation of Shear Zones in Granular
  Materials,} in \emph{Friction, Arching, Contact Dynamics}, edited by D.~E.
  Wolf, and P.~Grassberger, World Scientific, Singapore, 1997, pp. 233--247.

\bibitem[Az\'ema et~al.(2008)]{Azema2008}
E.~Az\'ema, F.~Radjai, and G.~Saussine, \emph{Mechanics of Materials}
  \textbf{accepted} (2008).

\end{thebibliography}

%%%%%%%%%%%%%%%%%%%%%%%%%%%%%%%%%%%%%%%%%%%
%% Just a reminder that you may have to run bibtex
%% All of it up to \end{document} can be removed
%% if you don't like the warning.
%%%%%%%%%%%%%%%%%%%%%%%%%%%%%%%%%%%%%%%%%%%
\IfFileExists{\jobname.bbl}{}
 {\typeout{}
  \typeout{******************************************}
  \typeout{** Please run "bibtex \jobname" to optain}
  \typeout{** the bibliography and then re-run LaTeX}
  \typeout{** twice to fix the references!}
  \typeout{******************************************}
  \typeout{}
 }

\end{document}